\documentclass[a4paper]{jpconf}
\usepackage{graphicx}
\usepackage{hyperref}
\begin{document}
\title{Scalar triplet on a domain wall}

\author{Vakhid A. Gani$^{1,2}$, Mariya A. Lizunova$^{1,2}$, Roman V. Radomskiy$^1$}

\address{$^1$National Research Nuclear University MEPhI (Moscow Engineering Physics Institute), Kashirskoe highway 31, Moscow, 115409, Russia}
\address{$^2$National Research Centre Kurchatov Institute, Institute for Theoretical and Experimental Physics, Bolshaya Cheremushkinskaya str. 25, Moscow, 117218, Russia}

\ead{vakhid.gani@gmail.com}

\begin{abstract}
We consider a model with a real scalar field with polynomial self-interaction of the fourth degree and a coupled scalar triplet. We demonstrate that there is an exact analytic solution in the form of a domain wall with a localised configuration of the scalar triplet coupled to the wall. We study some properties of this solution.
\end{abstract}

\section{Introduction}

Non-Abelian moduli fields are widely discussed in the literature, in particular, in the context of non-Abelian strings~\cite{Shifman01}-\cite{gorsky}. A number of theories allows strings with non-Abelian moduli both in supersymmetric and in non-supersymmetric models.

Recently several simple models were suggested that allow the existence of topological defects (Abrikosov-Nielsen-Olesen string or domain walls) which produce non-Abelian degrees of freedom on their world sheets~\cite{Shifman02,kur}.

We consider a model with a global non-Abelian symmetry $G$ which allows topological defects (strings, vertices, domain walls) to exist. These defects break the global symmetry $G$ to a subgroup $H$ (also global). Then, in addition to translational modes, there will be $\dim G-\dim H$ orientational modes. These moduli fields are governed by a sigma-model with $G/H$ internal space. As an example, see the recent publication~\cite{Shifman02} that introduced a simple model supporting domain walls. A domain wall breaks the global symmetry, and, as a result, there appear two orientational moduli in addition to a translational modulus. Besides that, in~\cite{kur} a spin-orbit interaction in the bulk was added that entangles the translational and orientational degrees of freedom.

Most of the main results of~\cite{kur} have been obtained numerically. We, in turn, investigate the same model mostly analytically. We obtained an exact static solution that can be seen as a domain wall with an attached lump of the $\chi$ field and examined some of its properties. We also study the small  excitation spectrum of the bare domain wall, with the field $\chi\equiv 0$, and confirm that this domain wall is unstable when the model parameters are chosen as in~\cite{kur}.

\section{The model and the exact solution}

We work with a real scalar field $\varphi$ and a real triplet of scalars $\chi^i$ in (3+1)-dimensional space-time. The space and time coordinates are $(x,y,z)$ and $t$, respectively. The dynamics of the system is governed by the Lagrangian
\begin{equation}\label{eq:Lagrangian}
{\cal L}=\frac{1}{2}\partial_\mu\varphi\partial^\mu\varphi+\frac{1}{2}\partial_\mu\chi^i\partial^\mu\chi^i-\lambda(\varphi^2-v^2)^2-\gamma\left[(\varphi^2-\mu^2)\chi^i\chi^i+\beta(\chi^i\chi^i)^2\right],
\end{equation}
where $\mu^2<v^2$, $\beta$, $\lambda$, and $\gamma$ are positive constants, and $i$ runs from 1 to 3.
The fields $\varphi$ and $\chi^i$ interact due to the term $-\gamma(\varphi^2-\mu^2)\chi^i\chi^i$ in the Lagrangian.

We further consider static domain-wall-like configurations. If the wall is orthogonal to the $z$ axis, the fields depend only on $z$,
\begin{equation}\label{eq:case}
\varphi=\varphi(z), \quad \chi^i=\chi^i(z),
\end{equation}
and the equations of motion for the fields $\varphi(z)$ and $\chi^i(z)$ that follow from the Lagrangian (\ref{eq:Lagrangian}) become
\begin{equation}\label{eq:eqmo1}
\varphi^{\prime\prime}=4\lambda\varphi(\varphi^2-v^2)+2\gamma(\chi^j\chi^j)\varphi,\quad
\chi^{i\prime\prime}=2\gamma(\varphi^2-\mu^2)\chi^i+4\beta\gamma(\chi^j\chi^j)\chi^i.
\end{equation}
Here the prime stands for the derivative with respect to $z$. Since the considered model (and, in particular, the system (\ref{eq:eqmo1}))
allows arbitrary orientation of the field $\chi^i$ in the internal space, we choose
\begin{equation}
\chi^i=\chi(z)\left(
\begin{array}{c}
0\\
0\\
1\\
\end{array}
\right).
\end{equation}
Then the system (\ref{eq:eqmo1}) takes the form:
\begin{equation}\label{eq:eqmo_main}
\varphi^{\prime\prime}=4\lambda\varphi(\varphi^2-v^2)+2\gamma\chi^2\varphi,\quad
\chi^{\prime\prime}=2\gamma\chi(\varphi^2-\mu^2)+4\beta\gamma\chi^3.
\end{equation}
To guess the form of the analytic solution, one has to look at the Lagrangian (\ref{eq:Lagrangian}).
If $\gamma=0$ then the fields do not interact, and the free field $\varphi(z)$ is described by the Lagrangian of the $\lambda\varphi^4$ model,
with one of the static solutions of the equation of motion being the well-known $\varphi^4$ kink~\cite{aek01}:
\begin{equation}\label{eq:kink}
\varphi(z)=\varphi_{\mbox {\scriptsize k}} (z)=v\tanh\left(\frac{m_{\varphi}}{2}z\right),
\end{equation}
where $m_{\varphi}=\sqrt{8\lambda v^2}$. Now, we assume that when we turn on the interaction between the fields in (\ref{eq:Lagrangian}),
$\varphi(z)$ holds its general form (\ref{eq:kink}) with the possible scale dilation~\cite{lensky}:
\begin{equation}\label{eq:phi}
\varphi=v\tanh\alpha z.
\end{equation}
When this ansatz is substituted into the first equation of the system (\ref{eq:eqmo_main}), we obtain for $\chi(z)$:
\begin{equation}\label{eq:chi}
\chi=\frac{A}{\cosh\alpha z}.
\end{equation}
Here the constant $A$ is related to $\alpha$ and the other model parameters by
\begin{equation}\label{eq:A}
A^2=\frac{2\lambda v^2-\alpha^2}{\gamma}.
\end{equation}
Further substituting (\ref{eq:phi}) and (\ref{eq:chi}) in the second equation of the system (\ref{eq:eqmo_main}) and equating coefficients in front of $\cosh^{-1}\alpha z$ and $\cosh^{-3}\alpha z$, we obtain two more relations between the constants:
\begin{equation}
\alpha^2=2\gamma(v^2-\mu^2),
\quad
\alpha^2=\gamma(v^2-2\beta A^2).
\end{equation}

Equations~(\ref{eq:phi}) and (\ref{eq:chi}), along with the constraints on the model parameters, are the static solution of the system (\ref{eq:eqmo_main}). Let us now discuss some features of this solution.

\begin{itemize}

\item
From equation~(\ref{eq:A}) it follows that $A=0$ at $\alpha=\sqrt{2\lambda v^2}$, which corresponds to the ``bare'' kink (\ref{eq:kink}). Besides that, the limit $\gamma\to 0$ can be taken only simultaneously with $\alpha\to\sqrt{2\lambda v^2}$.

\item
The spatial scales of the fields $\varphi$ and $\chi$ are the same, both of the order of $\alpha^{-1}$.

\item
The asymptotics of the fields at $z\to-\infty$ are
\begin{equation}
\varphi\approx-v+2v\:e^{2\alpha z}, \quad \chi\approx 2A\:e^{\alpha z}.
\end{equation}
Notice that these expressions coincide with those obtained in~\cite{kur}. The asymptotics of the fields at $z\to+\infty$ are
\begin{equation}
\varphi\approx v-2v\:e^{-2\alpha z}, \quad \chi\approx 2A\:e^{-\alpha z}.
\end{equation}

\item
On the $(\varphi,\chi)$ plane the solution (\ref{eq:phi}), (\ref{eq:chi}) is a semi-ellipse with semi-axes $v$ and $A$ (we assume that $A>0$). The ellipse becomes a line segment at $\alpha=\sqrt{2\lambda v^2}$.

\item
The energy of the solution (\ref{eq:phi}), (\ref{eq:chi}) or, more strictly, the tension of the corresponding plane structure in the three-dimensional space $(x,y,z)$,  is given by
\begin{equation}
E=\int\limits_{-\infty}^{+\infty}\left\{\frac{1}{2}\left(\frac{d\varphi}{dz}\right)^2+\frac{1}{2}\left(\frac{d\chi}{dz}\right)^2+\lambda(\varphi^2-v^2)^2+\gamma\left[(\varphi^2-\mu^2)\chi^2+\beta\chi^4\right]\right\}dz,
\end{equation}
and is equal to
\begin{equation}
E_{\mbox{\scriptsize sol}}=\frac{4A^4\beta\gamma+2v^2(\alpha^2+2\lambda v^2)+A^2(\alpha^2+2\gamma(v^2-3\mu^2))}{3\alpha}.
\end{equation}

When $A=0$ (and  $\alpha^2=2\lambda v^2$), this gives just the energy of the kink (\ref{eq:kink}):

$E_{\mbox{\scriptsize k}}=\displaystyle\frac{4\alpha v^2}{3}=\frac{2m_\varphi v^2}{3}$.

\end{itemize}

\section{A remark on the linear stability of the solution $\varphi_{\mbox{\scriptsize k}}(z)$, $\chi\equiv 0$}

It is easy to see that $\varphi=\varphi_{\scriptsize \mbox{k}}(z)$, $\chi\equiv 0$ satisfies the equations of motion following from the Lagrangian (\ref{eq:Lagrangian}).
Now we will investigate the linear stability of this configuration. To do this, we insert $\varphi=\varphi_{\scriptsize \mbox{k}}(z)$, $\chi=0+\delta\chi(t,z)$, where $\delta\chi(t,z)=\eta(z)\cos\omega t$, in the equations of motion, and get a time-independent Schr\"odinger-like spectral problem with the spectral parameter $\omega^2$:
\begin{equation}
-8\lambda v^2\eta_{\zeta\zeta}+2\gamma\left[v^2\tanh^2\left(\frac{\zeta}{2}\right)-\mu^2\right]\eta=\omega^2\eta,
\end{equation}
here $\zeta=m_\varphi z$, and $\eta=\eta(\zeta)$. The eigenvalues of $\omega$ are:
\begin{equation}
\omega_n^2=m_\varphi^2\:\varepsilon_n,\quad \varepsilon_n=\left(1-\frac{\mu^2}{v^2}\right)\frac{\gamma}{4\lambda}-\frac{1}{16}\left(-1-2n+\sqrt{1+\frac{4\gamma}{\lambda}}\right)^2.
\end{equation}
The discrete part of the spectrum is defined by $n<s=\displaystyle\frac{1}{2}\left(-1+\sqrt{1+\frac{4\gamma}{\lambda}}\right)$,
with $n$ starting from $n=0$. In particular, with the choice of the model parameters as in~\cite{kur}, we obtain:
\begin{equation}
s\simeq 2.37228,
\end{equation}
i.e.\ there are three levels in the discrete part of the spectrum:
\begin{equation}
\epsilon_0=-0.0119297,\quad
\epsilon_1=0.924211,\quad
\epsilon_2=1.36035.
\end{equation}
The presence of the negative value $\epsilon_0<0$ means $\omega_0^2<0$, i.e.\ the configuration $\varphi=\varphi_{\scriptsize \mbox{k}}(z)$, $\chi\equiv 0$ is unstable.
Notice that our value of $\epsilon_0$ differs from that reported in~\cite{kur}.

\section{Conclusion}

We have considered a model with a real scalar field and a coupled scalar triplet with the Lagrangian (\ref{eq:Lagrangian}). We found an exact analytic solution in the form of a domain wall with the coupled scalar triplet localized on the wall. We discussed some properties of this solution and also analysed the linear stability of the bare domain wall, and analytically showed that this bare domain wall solution is unstable.

\section*{Acknowledgments}

This work was supported in part by the Russian Federation Government under grant No.~NSh-3830.2014.2. V.~A.~Gani acknowledges support of the Ministry of Education and Science of the Russian Federation, Project No.~3.472.2014/K. M.~A.~Lizunova thanks the ITEP support grant for junior researchers and gratefully acknowledges financial support from the Dynasty Foundation.

This work was performed within the framework of the Center of Fundamental Research and Particle Physics supported by MEPhI Academic Excellence Project (contract No.~02.a03.21.0005, 27.08.2013).

\end{document}